\documentclass[journal=jpclcd, manuscript=letter]{achemso}
\setkeys{acs}{usetitle = true, etalmode = truncate, maxauthors = 55}

\usepackage{achemso}
\usepackage{graphicx} %
\usepackage{dcolumn}%
\usepackage{bm}%
\usepackage{booktabs}
\usepackage{textcomp}
\usepackage{makecell}
\usepackage{hhline}
\usepackage[usestackEOL]{stackengine}
\usepackage[version=3]{mhchem}
\usepackage{amsmath}
\usepackage{amsfonts}
\usepackage{braket}

\usepackage{multirow}

\usepackage[caption=false]{subfig}
\usepackage{tikz}

\title{Panoply of Ni-Doping-Induced Reconstructions, Electronic Phases, and Ferroelectricity in 1T-MoS$_2$}
\author{Rijan Karkee} 
\affiliation{Department of Physics, University of California, Merced, CA 95343}

\author{David A. Strubbe} 
\email{dstrubbe@ucmerced.edu}
\affiliation{Department of Physics, University of California, Merced, CA 95343}

\SectionNumbersOn
\begin{document}

\begin{tocentry}
    \includegraphics[scale=0.24]{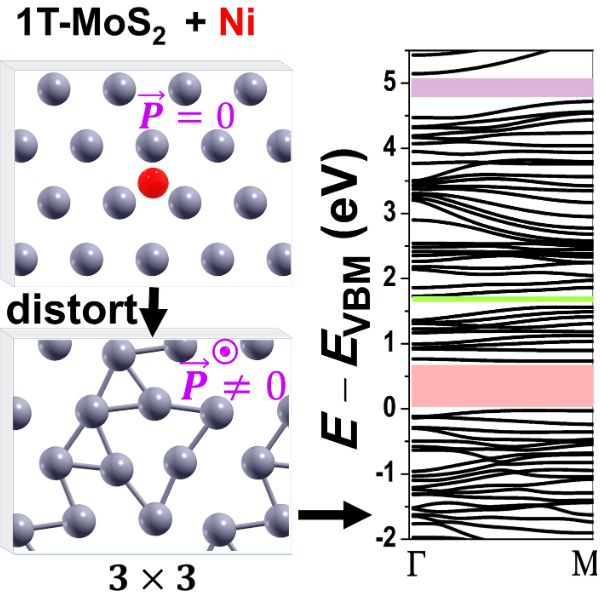}
\end{tocentry}

\begin{abstract}\label{Abstract}
The distorted phases of monolayer 1T-MoS$_2$ have distinct electronic properties, with potential applications in optoelectronics, catalysis, and batteries. We theoretically investigate the use of Ni-doping to generate distorted 1T phases, and find not only the ones usually reported but also two further phases (${3} \times {3}$ and ${4} \times {4}$), depending on the concentration and the substitutional or adatom doping site. Corresponding pristine phases are stable after dopant removal, which might offer a potential route to experimental synthesis. We find large ferroelectric polarizations, most notably in ${3} \times {3}$ which – compared to the recently measured 1T$''$ – has 100 times greater ferroelectric polarization, a lower energy, and a larger bandgap. Doped phases include exotic multiferroic semimetals, ferromagnetic polar metals, and improper ferroelectrics with only in-plane polarization switchable. The pristine phases have unusual multiple gaps in the conduction bands, with possible applications for intermediate band solar cells, transparent conductors, and nonlinear optics.

\end{abstract}
The crystal structure of MoS$_2$ has several polymorphs --
2H and 3R in bulk,
and in monolayers 1H and 1T -- which differ by coordination around Mo: 1H has trigonal prismatic coordination in common with 2H and 3R, and 1T has octahedral bonding. 1T is higher in energy, and so most studies of monolayer MoS$_2$ have focused on the 1H phase.
2H-MoS$_2$ has shown promise in lubrication, hydrodesulfurization, and optoelectronics. \cite{Mos2_application1,Mos2_application2,Mos2_application3,Mos2_application4}
Theoretical and experimental work has shown that doping with Ni and other transition metals in 1H- and 2H-MoS$_2$ can improve catalytic reactivity \cite{doped_apl1,doped_apl2,solar_hydro_pro}, lubrication and sliding \cite{appl_tribology,lubricant_by_David,Karkee2020,Guerrero_sliding,Acikgoz_2023} and gas sensing \cite{CO_and_NO_ads,doped_apl4}.
1T-MoS$_2$ and its well-known distortions to reconstructed phases \cite{1T_phase_review} have recently been shown to have a variety of exciting properties such as topological bands,\cite{1T_topological_insulator} high nonlinear optical susceptibility,\cite{1T_prime_3_exp} and ferroelectricity \cite{polarization_exp}.
However, much less is known about transition-metal doping in 1T-MoS$_2$ (as demonstrated experimentally with Ni intercalation \cite{Attanayake} and recently substitution for Mo \cite{Small_1T_Ni,1T_from_exp})
and its potential for modulating structural and electronic properties -- in particular, how transition-metal doping can interact with these distortions.

1T-MoS$_2$ shows metallic behavior, uniquely among the polytypes of MoS$_2$, which are otherwise semiconducting.
Applications of 1T-MoS$_2$ have been explored such as catalysis of the hydrogen evolution reaction \cite{1T_HER}, supercapacitors and batteries \cite{as_electrode_mos2,1T_batteries} and photocatalysis \cite{1T_photocatalysis}.
Theory \cite{1T_phase_review,1T_phase_recent} and experiments show 1T is unstable and instead is found in more stable distorted phases such as 1T$'$ (${2} \times 1$), 1T$''$ (${2}\times {2}$), 1T$'''$ ($ \sqrt{3} \times  \sqrt{3}$), $\sqrt{3} \times 1$, or $2\sqrt{3} \times 2\sqrt{3}$ \cite{1T_phase_more,1T_other_names,1T_phase_new_str}. (We will mostly use the reconstruction notation $N_a \times N_b$ for clarity.) 
These distorted 1T phases show interesting properties: 1T$'$ has been calculated to be a topological insulator \cite{1T_topological_insulator}, 1T$'''$ has been predicted to be a 2D ferroelectric material \cite{1T_ferroelectric} and a recent experiment showed a superior hydrogen evolution reaction activity compared to 2H-MoS$_2$ \cite{1T_triple_prime_exp}, and 1T$''$ was recently demonstrated experimentally to be ferroelectric.\cite{polarization_exp} 
Moreover, the different stackings of bulk 1T$'$ has been found to result in topological insulators or nodal-line semimetals,\cite{stackings} indicating yet another degree of freedom to explore in this family of materials.
Reconstructed 1T phases have been synthesized typically via intercalated alkali metal ions in 2H-MoS$_2$: Li intercalation and exfoliation leads to formation of 1T$'$ \cite{1T_prime_1_exp} and 1T$''$ \cite{1T_prime_2_exp}, and synthesis of KMoS$_2$ and deintercalation of K leads to 1T$'$ and 1T$'''$.\cite{1T_prime_3_exp} Intercalation with transition metals Cu, Ni, and Co has also been used to synthesize 1T phases \cite{Cu_intercalation, Attanayake}.
Calculations showed that adsorption of Li on 1H \cite{1T_doping_distortion_Li}, Cu adsorption on 2H \cite{Cu_1T}, and charge doping of 2H \cite{1T_phase_data_match} or 1T\cite{1T_doping_distortion} can also lead to 1T$'$. Other non-doping-related synthesis methods include laser writing on 2H to produce 1T$'$ regions \cite{Mine_laser}, tensile strain on the corresponding 2H-MoTe$_2$ to produce 1T$'$-MoTe$_2$,\cite{1T_external_stimuli} and direct synthesis methods for 1T-MoS$_2$ in vapor and liquid phases.\cite{1T_other_names}
The variety of 1T-derived phases of MoS$_2$, the range of remarkable properties they show, and the use of intercalation, adsorption, and charge doping for their synthesis, together suggest the possibility of using other dopants to further explore the potential phase space for 1T reconstructions, searching for novel properties as well as ways to tune or optimize known properties of 1T phases.

\begin{figure}[h]
	\includegraphics[scale=0.7]{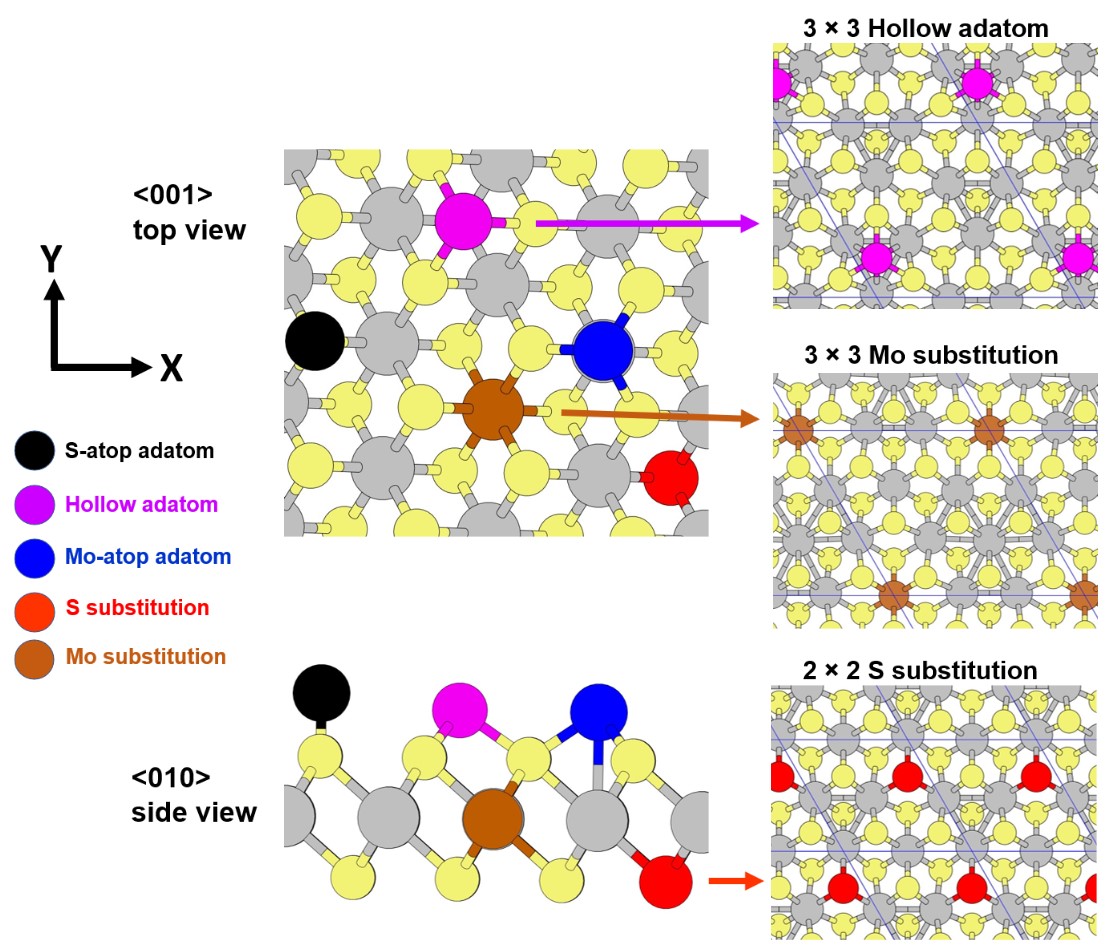}
	\caption{\small{Left: Different doping sites for Ni in 1T-MoS$_2$, with definition of $x$ and $y$ axes.
	Right: Example reconstructions for different doping sites and concentrations.}}
	\label{fig:structures}
\end{figure}

In this paper, we perform systematic first-principles calculations to investigate the effect of Ni-doping on 1T-MoS$_2$ at various concentrations, including substitutions for Mo and S as well as adsorption on different sites.
We consider the most probable sites for dopants in monolayer MoS$_2$, including substitutions and adatoms (same as in 1H \cite{Karkee2020}), as shown in Fig. \ref{fig:structures}. (While adsorption may be considered a distinct process from doping, we classify it together with substitution here due to its connection with bulk intercalation and potential charge doping.\cite{enrique}) The bridge adatom site relaxes to hollow and is not considered further. We study $2\times2$, $3\times3$, and $4\times4$ supercells, each with one Ni dopant, corresponding to different doping concentrations. While an ordered doping scheme might not be necessary, it has been frequently found that high doping concentrations lead to ordered phases.\cite{Na_phases,zero_valent}

	We compare the thermodynamic favorability via the doping formation energy, a function of the chemical potentials during growth\cite{enrique} (SI Sec. 2). Our reference is the most stable distorted phase, 1T$'$ ($2\times1$), which is 0.26 eV per MoS$_2$ unit lower in energy than 1T.
		The doping formation energies for Mo or S substitution and different adatom sites
		are shown in Figs. S1-S3. Hollow is the most favorable adatom site,\cite{1T_ringcenter} though Mo-atop is favored for 1H \cite{Karkee2020}. 
		Although these energies are positive, Ni doping is energetically favorable from undistorted 1T in nearly all cases due to lowering of energy by reconstructions. Substitutional doping formation energies are more favorable for 1T$'$ than 1H.\cite{Karkee2020} Doping lowers energy differences between 1H and 1T compared to the pristine $\Delta E = 0.85$ eV per MoS$_2$, but 1H remains favored (Fig. S4). Mo substitution at high concentration (8\%) can bring the difference down to 0.23 eV per MoS$_2$ unit. The $T=0$ phase diagram as a function of chemical potential\cite{enrique} (Fig. S5(a)) shows that in equilibrium the hollow site is most compatible with stability of 1T or 1T$'$, but Mo and S substitution are also accessible under some conditions. We note that our previous work showed that Ni-doped MoS$_2$ is always above the convex hull at $T=0$ compared to other compounds of Ni, Mo, and S.\cite{enrique} Despite this thermodynamic unfavorability, abundant experimental evidence has been found of Ni incorporation into 2H-MoS$_2$\cite{Ni_intercalation_oxidation_state,Ni_2H_3R}, and more recently, 1T-MoS$_2$.\cite{Attanayake,Small_1T_Ni,1T_from_exp} Approaches can involve non-equilibrium conditions such as hydrothermal methods for Mo substitution, or creation of sulfur vacancies that can be filled by transition metals for S substitution.\cite{Liu_S_vacancy,thomas2023substitutional}
	  Doping is found to induce different reconstructions, with various patterns of Mo-atom clustering depending on supercell size and doping scheme (Fig. \ref{fig:structures}). The DFT relaxation trajectories show that the distortions are driven by bond length changes when Ni is introduced to the structure, which not only break the symmetry but also define the local geometry. Mo atoms whose bonds to S are stretched are induced to cluster, which can in turn stretch other Mo-S bonds and induce further clustering. 
	  As in 1H-MoS$_2$ \cite{Karkee2020}, we find Ni-S bonds in Mo-substitution are shorter (2.25 \rm\AA) and Ni-Mo bonds in S-substitution are longer (2.51 \rm\AA), compared to Mo-S bonds. In Mo substitution, Ni pulls S closer, weakening neighboring Mo-S bonds and making the Mo atoms cluster. In S substitution, Ni pushes Mo farther away, causing now-closer Mo atoms to cluster. In S-atop, Ni weakens bonds from that S to Mo neighbors, causing clustering around Ni. In Mo-atop, Ni pulls S atoms closer, causing clustering away from Ni of Mo atoms with stretched bonds. In hollow, Ni pushes Mo away, causing clustering away from Ni.

\begin{table}[h!tbp]
  \centering

  \caption{Different pristine reconstructions resulting after removal or replacement of Ni atom in different supercells of Ni-doped 1T.}
    \begin{tabular}{|c|c|c|c|}
    \hline
    \textbf{Doping type} & \textbf{$2\times 2$} & \textbf{$3\times 3$} & \textbf{$4\times 4$} \\\hline \hline
    Mo subs. & $2 \times 2$ & $3 \times 3$ & $4 \times 4$ \\
    \hline
        \multicolumn{1}{|c|}{S subs.} & \multicolumn{1}{c|}{$2 \times 2$} & \multicolumn{1}{c|}{$ \sqrt{3} \times \sqrt{3}$ 4B} & \multicolumn{1}{c|}{$2 \times 2$} \\
    \hline
        \multicolumn{1}{|c|}{Hollow} & \multicolumn{1}{c|}{$2 \times 2$} & \multicolumn{1}{c|}{$ \sqrt{3} \times  \sqrt{3}$ 2B} & \multicolumn{1}{c|}{$2 \times 2$} \\
    \hline
        \multicolumn{1}{|c|}{Mo-atop} & \multicolumn{1}{c|}{$2 \times 2$} & \multicolumn{1}{c|}{$ 3 \times 3$} & \multicolumn{1}{c|}{$2 \times 2$} \\
 \hline
   \multicolumn{1}{|c|}{S-atop} & \multicolumn{1}{c|}{$2 \times 2$} & \multicolumn{1}{c|}{$ \sqrt{3} \times  \sqrt{3}$ 2B} & \multicolumn{1}{c|}{$2 \times 2$} \\
   \hline
    \end{tabular}%
      \label{tab:1T_structure_relaxations}
\end{table}%

\begin{figure}[h]
			\includegraphics[scale=.5]{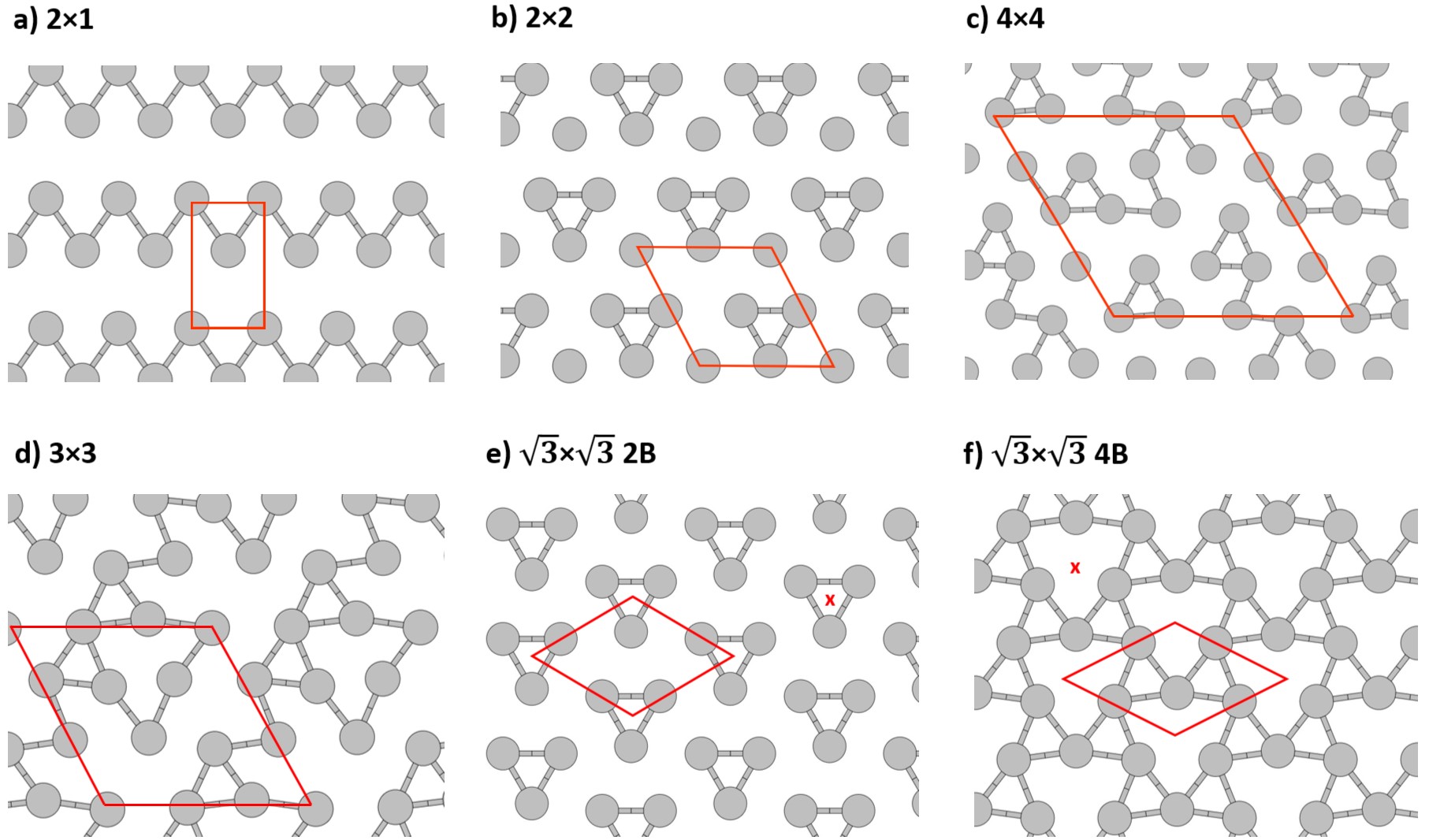}
			\caption{Mo clustering for different reconstructions. S atoms are not shown, for clarity. All Mo-Mo bonds shorter than 3.08 \AA, twice the covalent radius of Mo, \cite{covalent_radii} are shown. There are several inequivalent bond lengths in $3\times3$ and $4\times4$.
			The red $\times$ in (e) and (f) denotes the location of the C$_3$ axis. }
			\label{fig:pristine_clustering}
			
		\end{figure}

	We found that the doped structures can give rise to  reconstructions even of the pristine material: we reversed the doping by removing adatoms or restoring Mo or S atoms to Ni-substituted sites, and relaxed the resulting structures, finding stable distortions. This process of removal of adatoms is akin to the methods used to synthesize 1T phases via Li or K atoms \cite{1T_prime_Kdoping}, and undoing Ni substitution of Mo or S seems plausible via ion-exchange methods that have been used to interconvert different TMDs \cite{Chen_ionexchange} and produce 1T-MoS$_2$ nanoflowers.\cite{nanoflowers} Therefore the route used in our calculations may also constitute a viable method for experimental synthesis, though stabilization by epitaxy \cite{CHAE201965} or ligands \cite{ligands} are also possible strategies.
	Most of the structures relaxed to previously known phases ($2 \times 2$ or $ \sqrt{3} \times  \sqrt{3} $) \cite{1T_phase_data_match,1T_phase_more,1T_phase_recent,1T_phase_review} while a few reconstructions were not found in the literature (Table \ref{tab:1T_structure_relaxations}, Fig. \ref{fig:pristine_clustering}). Note that the supercell size constrains the possible reconstructions: e.g. $2\times2$ can form only from $2\times2$ or $4\times4$, and $ \sqrt{3} \times  \sqrt{3}$ only from $3 \times 3$. We did not find $2\times1$ but include it in further analysis for comparison. The properties of the pristine reconstructions are shown in Table \ref{tab:1T_structure_properties}. Lattice parameters and Mo-Mo distances are consistent with \citet{1T_phase_data_match} %
	We find there are two different $ \sqrt{3} \times  \sqrt{3}$ structures, with the same $C_{3v}$ symmetry (but different locations of the $C_3$ axis) and all 3 Mo atoms equivalent at the Wyckoff $c$ positions. They have different numbers of bonds from Mo to other Mo atoms, so we will distinguish them as 2B and 4B.
	The distinction between these structures does not seem to have been appreciated previously: $\sqrt{3} \times  \sqrt{3}$ in \citet{1T_phase_data_match} is 2B whereas in \citet{1T_prime_3_exp} it is 4B. %
    The phases not previously reported are obtained from $3\times3$ and $4\times4$ Mo substitution, and the energies are 0.24 eV and 0.15 eV  per  MoS$_2$ lower than 1T, respectively. Note that $3\times3$ is more stable than the experimentally established $2 \times 2$.\cite{polarization_exp} %
    The most stable structure seen is $\sqrt{3} \times \sqrt{3}$ 4B, which has been synthesized experimentally \cite{1T_prime_3_exp} with lattice parameter $a=b=$ 5.58 \rm\AA\space and Mo-Mo bond length 3.01 \rm\AA, close to our results of 5.61 \rm\AA\space and 3.02 \rm\AA.  
    Space groups and point groups of the structures are given in Table S3. We confirmed that similar results are obtained for the relative energies and structural parameters using PBE+$U$ \cite{Dudarev} or the PBE0 hybrid functional\cite{Ernzerhof1999,Adamo1999} (Table S4). 

\begin{figure}[h!]
		
			\centering{
			\begin{tikzpicture}
			\node [anchor=north west] (imgA) at (-0.15\linewidth,.58\linewidth){\includegraphics[width=0.325\linewidth]{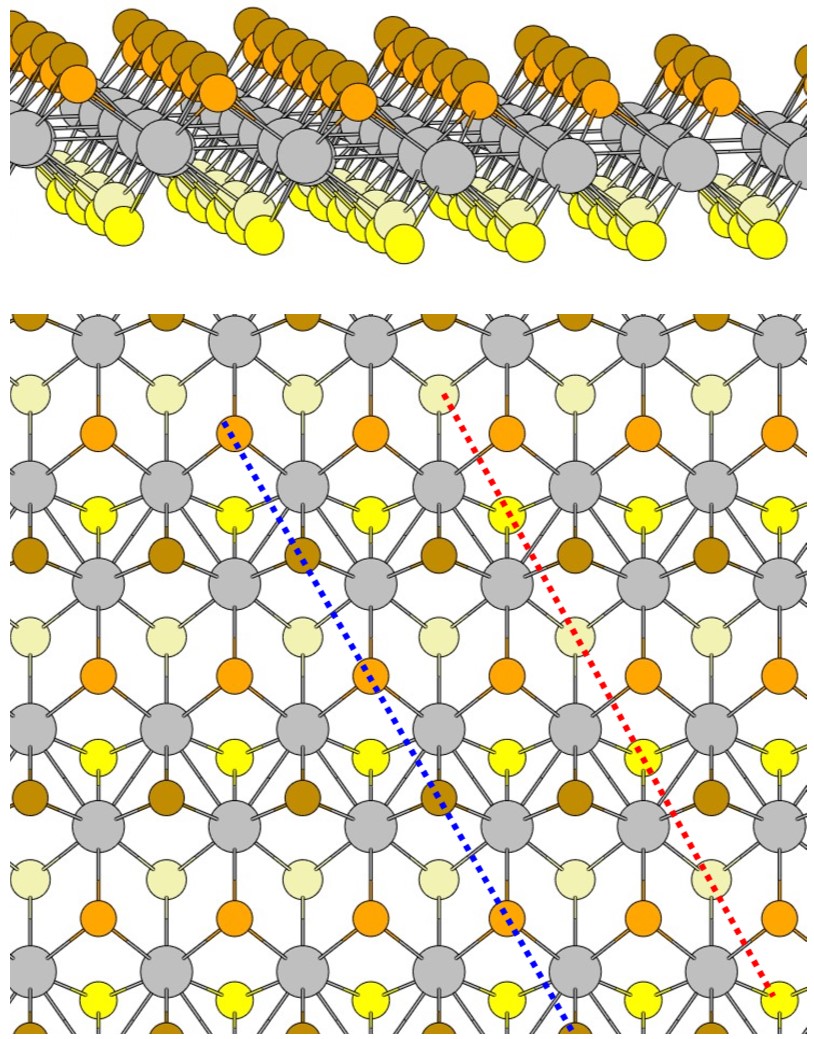}};
            \node [anchor=north west] (imgB) at (0.186\linewidth,.58\linewidth){\includegraphics[width=0.33\linewidth]{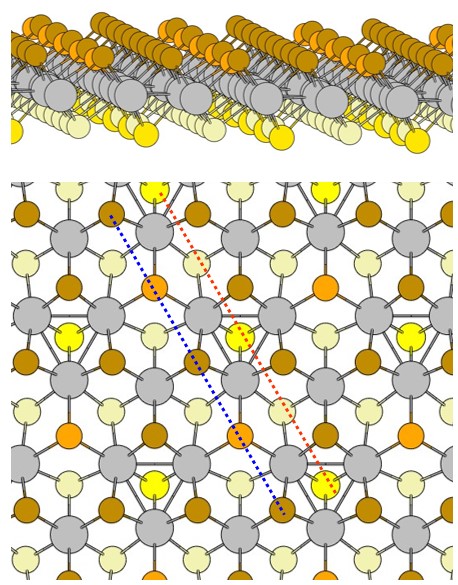}};
          \node [anchor=north west] (imgC) at (0.52\linewidth,.58\linewidth){\includegraphics[width=0.33\linewidth]{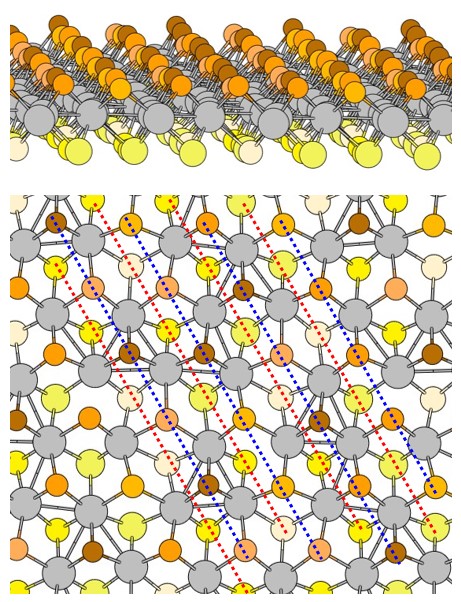}};
          
          	\node [anchor=north west] (imgD) at (-0.15\linewidth,.07\linewidth){\includegraphics[width=0.33\linewidth]{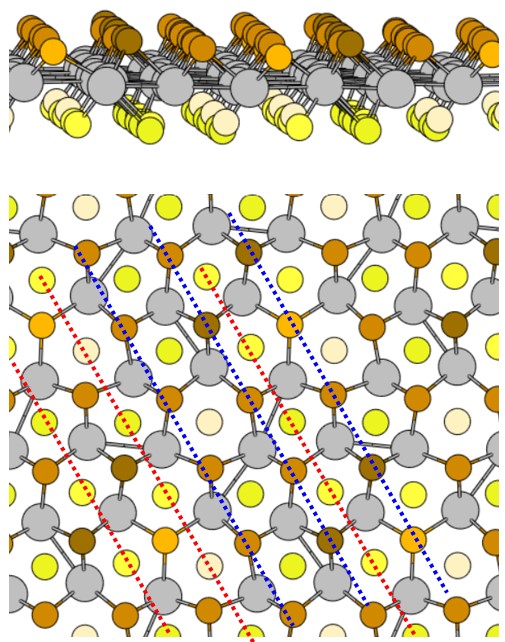}};
            \node [anchor=north west] (imgE) at (0.186\linewidth,.08\linewidth){\includegraphics[width=0.325\linewidth]{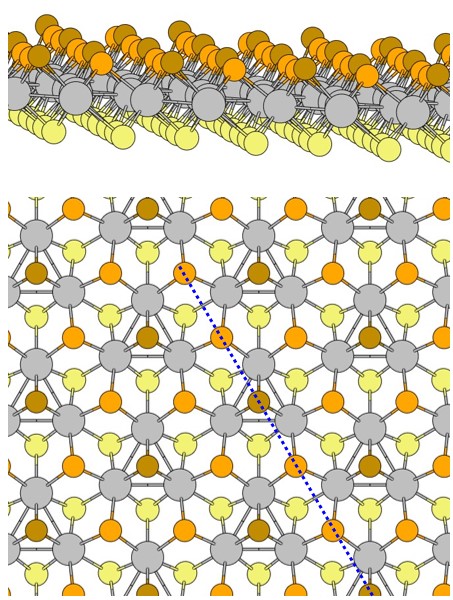}};
          \node [anchor=north west] (imgF) at (0.52\linewidth,.065\linewidth){\includegraphics[width=0.33\linewidth]{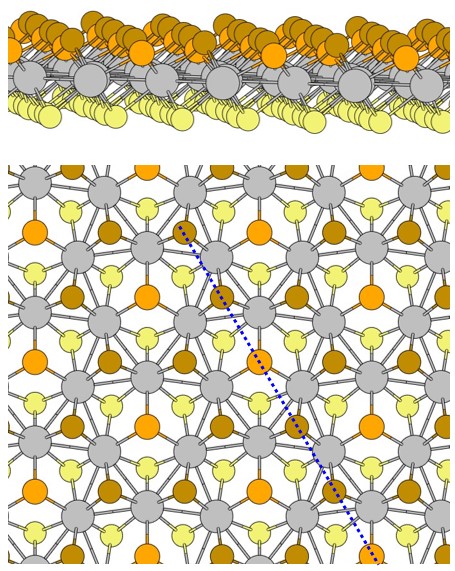}};
          
            \draw [anchor=north west] (-0.15\linewidth, .61\linewidth) node {(a) {\fontfamily{Arial}\selectfont \textbf{$2\times1$}}};
            \draw [anchor=north west] (0.19\linewidth, .61\linewidth) node {(b) {\fontfamily{Arial}\selectfont \textbf{$2\times2$}}};
           \draw [anchor=north west] (0.52\linewidth, .61\linewidth) node {(c) {\fontfamily{Arial}\selectfont \textbf{$4\times4$}}};
            \draw [anchor=north west] (-0.15\linewidth, .11\linewidth) node {(d) {\fontfamily{Arial}\selectfont \textbf{$3\times3$}}};
            \draw [anchor=north west] (0.19\linewidth, .11\linewidth) node {(e) {\fontfamily{Arial}\selectfont \textbf{$\sqrt{3} \times \sqrt{3}$ 2B}}};
           \draw [anchor=north west] (0.52\linewidth, .11\linewidth) node {(f) {\fontfamily{Arial}\selectfont \textbf{$\sqrt{3} \times \sqrt{3}$ 4B}}};           
           
            \end{tikzpicture}
			}
			
			\caption{Structures in side and top views, showing vertical corrugation of S atoms. Orange and yellow colors are used to represent S atoms on the top and bottom of the Mo-plane, respectively. Darker color indicates greater distance from the Mo-plane. Blue and red dotted lines indicate directions along which S atoms heights vary for top and bottom S atoms, respectively.
			}
		\label{fig:vertical_structures}
			
		\end{figure}

    These structures are distorted out of plane as well: S atoms vary in their $z$-height (Fig. \ref{fig:vertical_structures}), a symmetry-breaking which allows out-of-plane polarization. There has been little attention in distorted 1T to this $z$-variation \cite{1T_phase_data_match, 1T_phase_more,1T_phase_review,1T_prime_3_exp,1T_ferroelectric}, and only recently has it been connected to polarization in the $2\times2$ structure.\cite{polarization_exp}
    The variations for $3 \times 3$ are largest, and as shown below, it has the largest ferroelectric polarization. While the maximum displacements are similar on the two sides, there is a net polarization because the patterns are different, and there can be significant variation in dynamical charges among S atoms in 1T-MoS$_2$ phases.\cite{polarization_exp,Choi_2020} In $\sqrt{3}\times \sqrt{3}$ 2B and 4B, the $z$-height of S atoms varies on one side and is constant on the other, whereas in all other structures, we find variations on both sides. The pattern of $z$-variation in 2B and 4B is opposite: in 2B, lines of S atoms consist of pairs closer to the Mo plane separated by an atom farther; whereas in 4B there are pairs farther, separated by an atom closer. The variation on only one side in $\sqrt{3}\times \sqrt{3}$ is connected to the fact that these reconstructions only occur from an adatom or S substitution, which breaks the symmetry between the sides, but not from Mo substitution. We find Ni substituting S causes sites to move closer to the Mo plane, and Ni on S-atop and Mo-atop sites moves S farther from Mo. Ni on the hollow site moves some S atoms closer and some farther, on both sides.
    The S atoms centered in Mo triangular clusters are always farthest from the Mo plane.
    These vertical corrugations are consistent with previously calculated structures of $\sqrt{3}\times \sqrt{3}$ 4B in \citet{1T_prime_3_exp}, $2\times2$ in \citet{polarization_exp}, and $\sqrt{3}\times \sqrt{3}$ 2B (in plane \cite{1T_phase_data_match} and out of plane \cite{private_comm}). This surface non-uniformity could be interesting for surface chemistry or catalysis as it may tune reactivity:
    in 1T$'$ the reactivity for chemisorption is higher for S atoms closer to the Mo plane. \cite{S_plane_reactivity}

\begin{table}[htbp]
  \centering
    \begin{tabular}{cccccc}
    \multicolumn{5}{l}{\multirow{2}[2]{*}{a) }} \\
    \multicolumn{5}{r}{} \\
     \cline{1-6}
    \cline{1-6}
    \cline{1-6}
    \cline{1-6}
    
    \multicolumn{1}{|c|}{Structure} & \multicolumn{1}{c|}{$a, b$ (\rm\AA)} & \multicolumn{1}{c|}{$d$ (\rm\AA)} &
    \multicolumn{1}{c|}{$\Delta z$ (\rm\AA)} & \multicolumn{1}{c|}{$\Delta E$ (eV)} & \multicolumn{1}{c|}{$E_g$ (eV)} \\
    \Xhline{2\arrayrulewidth}
    \multicolumn{1}{|c|}{ $1\times1$} & \multicolumn{1}{r|} {3.18}     & \multicolumn{1}{r|}{3.18 } &
    \multicolumn{1}{r|}{0, 0} &
    \multicolumn{1}{r|}{0}     & \multicolumn{1}{r|}{0} \\
    \multicolumn{1}{|c|}{ $2 \times 1$} & \multicolumn{1}{r|} {5.67, 3.18}     & \multicolumn{1}{r|}{2.78} &
    \multicolumn{1}{r|}{0.39, 0.39} & \multicolumn{1}{r|}{-0.26}     & \multicolumn{1}{r|}{0.02} \\
    \multicolumn{1}{|c|}{$2 \times 2$} & \multicolumn{1}{r|}{6.43}    & \multicolumn{1}{r|}{2.78} &
    \multicolumn{1}{r|}{0.27, 0.25} & \multicolumn{1}{r|}{-0.21}    & \multicolumn{1}{r|}{0.08}\\
    \multicolumn{1}{|c|}{$3 \times 3$} & \multicolumn{1}{r|}{9.70}     & \multicolumn{1}{r|}{2.93} &
    \multicolumn{1}{r|}{0.47, 0.51} & \multicolumn{1}{r|}{-0.24}     & \multicolumn{1}{r|}{0.71} \\
    \multicolumn{1}{|c|}{$\sqrt{3} \times \sqrt{3}$ 2B} & \multicolumn{1}{r|}{5.64}     & \multicolumn{1}{r|}{2.84} &
    \multicolumn{1}{r|}{0, 0.17} & \multicolumn{1}{r|}{-0.19}    & \multicolumn{1}{r|}{0.59} \\
      \multicolumn{1}{|c|}{$\sqrt{3} \times \sqrt{3}$ 4B} & \multicolumn{1}{r|}{5.61}     & \multicolumn{1}{r|}{3.02} &
      \multicolumn{1}{r|}{0, 0.44} & \multicolumn{1}{r|}{-0.26}    & \multicolumn{1}{r|}{0.83} \\
      \multicolumn{1}{|c|}{$4 \times 4$} & \multicolumn{1}{r|}{13.01}     & \multicolumn{1}{r|}{2.78} &
      \multicolumn{1}{r|}{0.32, 0.31} & \multicolumn{1}{r|}{-0.15}    & \multicolumn{1}{r|}{0.25} \\

   \cline{1-6}
    \cline{1-6}
    \cline{1-6}
    \cline{1-6}
    
    \multicolumn{4}{l}{\multirow{2}[2]{*}{b)  }} \\
    \multicolumn{4}{r}{} \\
    \cline{1-4}
    \cline{1-4}
    \cline{1-4}
    \cline{1-4}
    \multicolumn{1}{|c|}{Structure}  & \multicolumn{1}{c|}{Bandgap (eV)} & \multicolumn{1}{c|}{$\Delta E_{\rm VBM}$ (eV)} & \multicolumn{1}{c|}{$N_{\rm bands}$} \\
    \cline{1-4}
    \cline{1-4}
    \cline{1-4}
    \cline{1-4}
     \multicolumn{1}{|c|}{} &      \multicolumn{1}{r|}{0.46} & \multicolumn{1}{r|}{1.63} & \multicolumn{1}{r|}{2}  \\
    \multicolumn{1}{|c|}{$1\times1$}       & \multicolumn{1}{r|}{0.58} & \multicolumn{1}{r|}{3.98} & \multicolumn{1}{r|}{2} \\
    \cline{1-4}
    \multicolumn{1}{|c|}{$2 \times 1$}      & \multicolumn{1}{r|}{0.30} & \multicolumn{1}{r|}{4.81} & \multicolumn{1}{r|}{8}     \\
  
     \cline{1-4}
    \multicolumn{1}{|c|}{}     & \multicolumn{1}{r|}{0.19} & \multicolumn{1}{r|}{0.81} & \multicolumn{1}{r|}{2}  \\
    \multicolumn{1}{|c|}{$2 \times 2$}       & \multicolumn{1}{r|}{0.15} & \multicolumn{1}{r|}{1.92} & \multicolumn{1}{r|}{3}\\
    \multicolumn{1}{|c|}{}      & \multicolumn{1}{r|}{0.52} & \multicolumn{1}{r|}{4.52} & \multicolumn{1}{r|}{11}  \\
    \cline{1-4}
    \multicolumn{1}{|c|}{}      & \multicolumn{1}{r|}{0.10} & \multicolumn{1}{r|}{1.63} & \multicolumn{1}{r|}{9}  \\
    \multicolumn{1}{|c|}{$3 \times 3$}      & \multicolumn{1}{r|}{0.42} & \multicolumn{1}{r|}{4.72} & \multicolumn{1}{r|}{27}  \\

    \cline{1-4}
    \multicolumn{1}{|c|}{}    & \multicolumn{1}{r|}{0.83} & \multicolumn{1}{r|}{1.20}   & \multicolumn{1}{r|}{3}  \\
    \multicolumn{1}{|c|}{$\sqrt{3} \times \sqrt{3}$ 2B}     & \multicolumn{1}{r|}{0.22} & \multicolumn{1}{r|}{2.54}  & \multicolumn{1}{r|}{3}  \\
    \multicolumn{1}{|c|}{}       & \multicolumn{1}{r|}{0.73} & \multicolumn{1}{r|}{4.40} & \multicolumn{1}{r|}{6} \\
    \cline{1-4}
    \multicolumn{1}{|c|}{}    & \multicolumn{1}{r|}{0.22} & \multicolumn{1}{r|}{1.70}  & \multicolumn{1}{r|}{3}  \\
     \multicolumn{1}{|c|}{$\sqrt{3} \times \sqrt{3}$ 4B}       & \multicolumn{1}{r|}{0.38} & \multicolumn{1}{r|}{4.89} & \multicolumn{1}{r|}{9}\\
     \cline{1-4}

    \multicolumn{1}{|c|}{$4 \times 4$}       & \multicolumn{1}{r|}{0.11} & \multicolumn{1}{r|}{1.84} & \multicolumn{1}{r|}{17} \\
    \multicolumn{1}{|c|}{}      & \multicolumn{1}{r|}{0.39} & \multicolumn{1}{r|}{4.69} & \multicolumn{1}{r|}{47}  \\
    \cline{1-4}
    \cline{1-4}
    \cline{1-4}
    \cline{1-4}
    \end{tabular}
  \caption{Structure, energy, and DFT bandstructure properties of pristine reconstructions of 1T.\textsuperscript{\emph{*}}}
\textsuperscript{\emph{*}} a) Structure, energy, and gap; $a,b$ are the lattice parameters ($a=b$ in most cases); $d$ is the shortest Mo-Mo distance, characterizing clustering; $\Delta z$ gives the range of vertical positions of S atoms, for each of the two sides; $\Delta E$ is the energy difference per MoS$_2$ with respect to undistorted 1T. b) Conduction-band gaps: $\Delta E_{\rm VBM}$ is the energy difference of the bottom of the gap from the valence band maximum (VBM); $N_{\rm bands}$ is the number of bands from the previous gap up to this gap.
\label{tab:1T_structure_properties}
\end{table}

The pristine reconstructions result from relaxation of a structure whose symmetry was broken by the presence of the Ni dopants, which indicates a minimum level of stability. The dynamical stability of all reconstructions was further investigated by calculating phonon bandstructures, as shown in Fig. S13. We found structures $2\times1$, $3\times3$, and $\sqrt{3}\times\sqrt{3}$ 4B to be dynamically stable. However, in several cases instabilities were found -- real in the case of $\sqrt{3}\times\sqrt{3}$ 2B and $3\times3$, but attributed to numerical issues \cite{Pallikara_2022} in the case of $2\times1$, $2\times2$ and $4\times4$. The instabilities were investigated by displacing the atomic coordinates in the direction of the imaginary-frequency displacement eigenvectors scaled by 0.01 \AA, and then relaxing the atomic structure and lattice.\cite{Pallikara_2022} The higher-energy $\sqrt{3}\times\sqrt{3}$ structure (2B) \cite{1T_phase_data_match} was found to be unstable, with imaginary frequencies at $\Gamma$, M and K. Relaxation after displacement by the imaginary mode at $\Gamma$ yielded $\sqrt{3}\times\sqrt{3}$ 4B. We nonetheless retained $\sqrt{3}\times\sqrt{3}$ 2B structure in our data set due to interesting properties such as a higher ferroelectric polarization (see below), and the possibility that it might be stabilized by temperature, epitaxy,\cite{CHAE201965} ligands,\cite{ligands} or other factors. Our originally obtained $3\times3$ structure (Fig. S14), which is 0.08 eV lower in energy than undistorted 1T, proved to be unstable, and relaxation after displacement led to the significantly more stable $3\times3$ structure analyzed in the text. In the case of $2\times2$, we found a small imaginary frequency at $\Gamma$ which corresponds to a flexural mode (Fig. S15), notoriously difficult to calculate accurately in DFT.\cite{Croy_2020} Relaxation after displacement returned to the original structure, showing that the instability is a numerical artifact, which is consistent with the realization in experiment of this phase.\cite{polarization_exp} $2\times1$ shows a similar flexural instability at $\Gamma$, which we did not investigate further, since it is included here only for comparison to the phases resulting from Ni-doping, and the existence of this 1T$'$ phase is well established experimentally\cite{1T_prime_1_exp,1T_prime_3_exp,Mine_laser} and theoretically.\cite{1T_prime_band,1T_phase_data_match} For $4\times4$, relaxation in a $2\times1$ supercell after displacement by imaginary modes corresponding to the M point led back to the original structure, indicating no true instability. These modes also have flexural character (Fig. S15). We conclude that the $2\times2$ and $4\times4$ structures are indeed dynamically stable.

\begin{figure}[h]
			\centering{
			\begin{tikzpicture}
			\node [anchor=north west] (imgA) at (-0.15\linewidth,.58\linewidth){\includegraphics[width=0.335\linewidth]{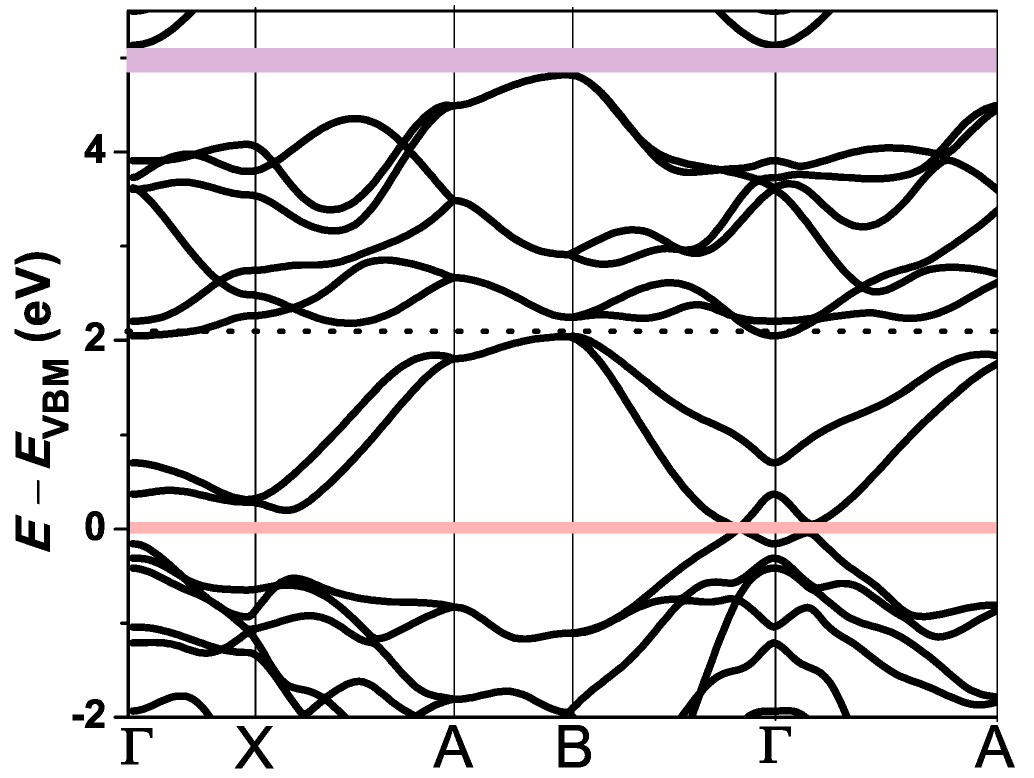}};
            \node [anchor=north west] (imgB) at (0.186\linewidth,.58\linewidth){\includegraphics[width=0.33\linewidth]{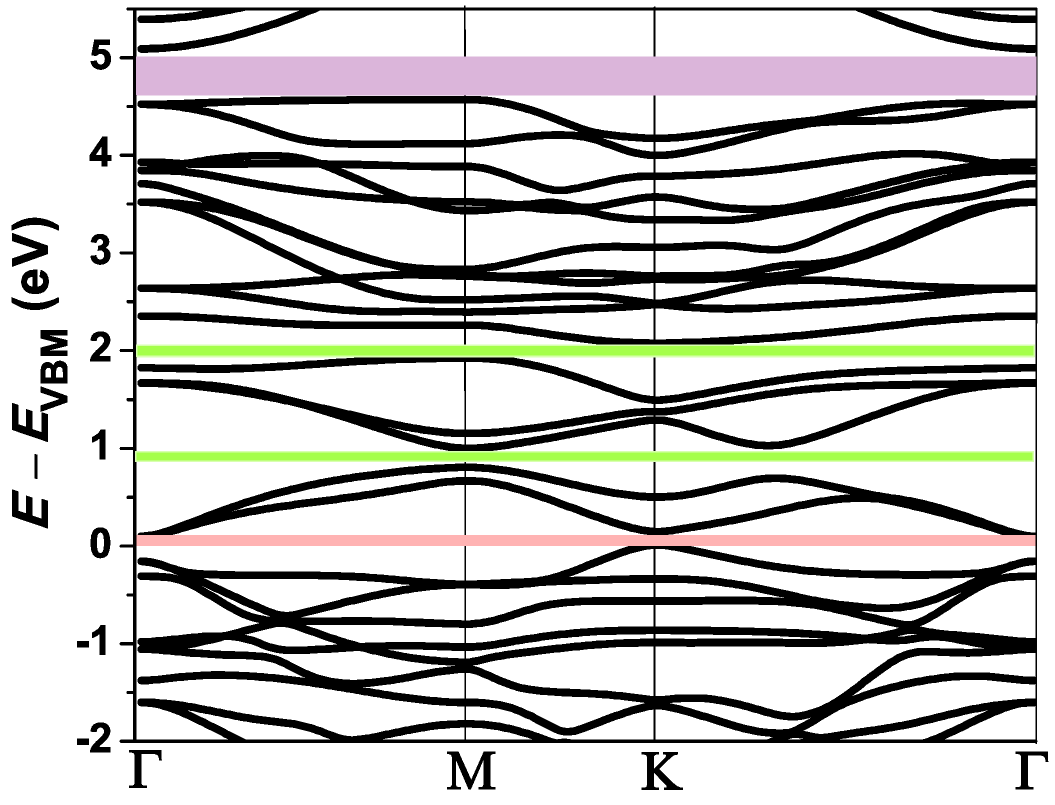}};
          \node [anchor=north west] (imgC) at (0.52\linewidth,.58\linewidth){\includegraphics[width=0.33\linewidth]{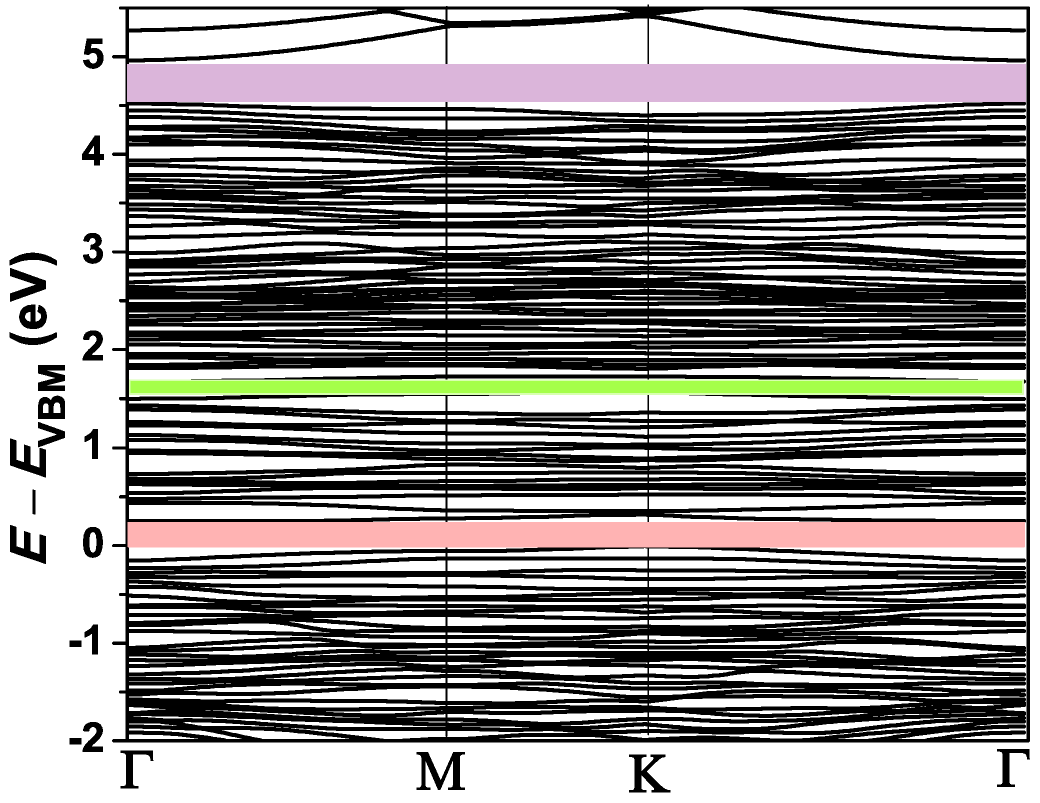}};
          
          	\node [anchor=north west] (imgD) at (-0.15\linewidth,.265\linewidth){\includegraphics[width=0.333\linewidth]{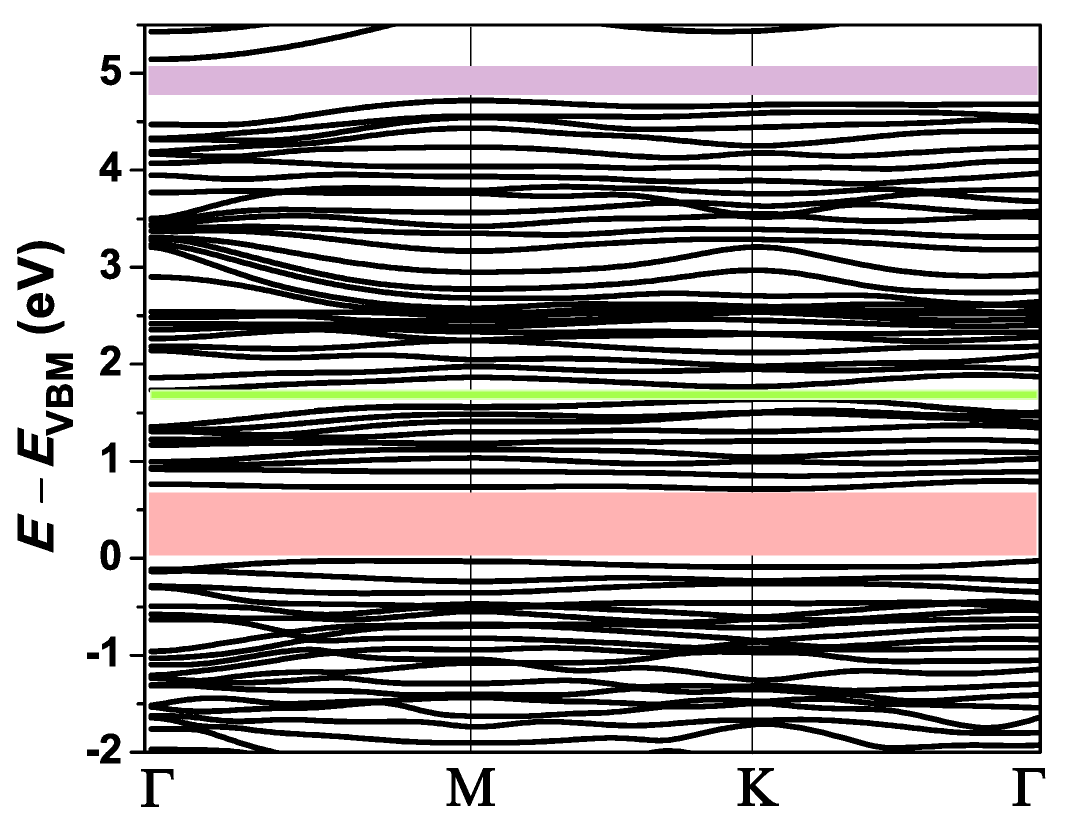}};
            \node [anchor=north west] (imgE) at (0.186\linewidth,.26\linewidth){\includegraphics[width=0.33\linewidth]{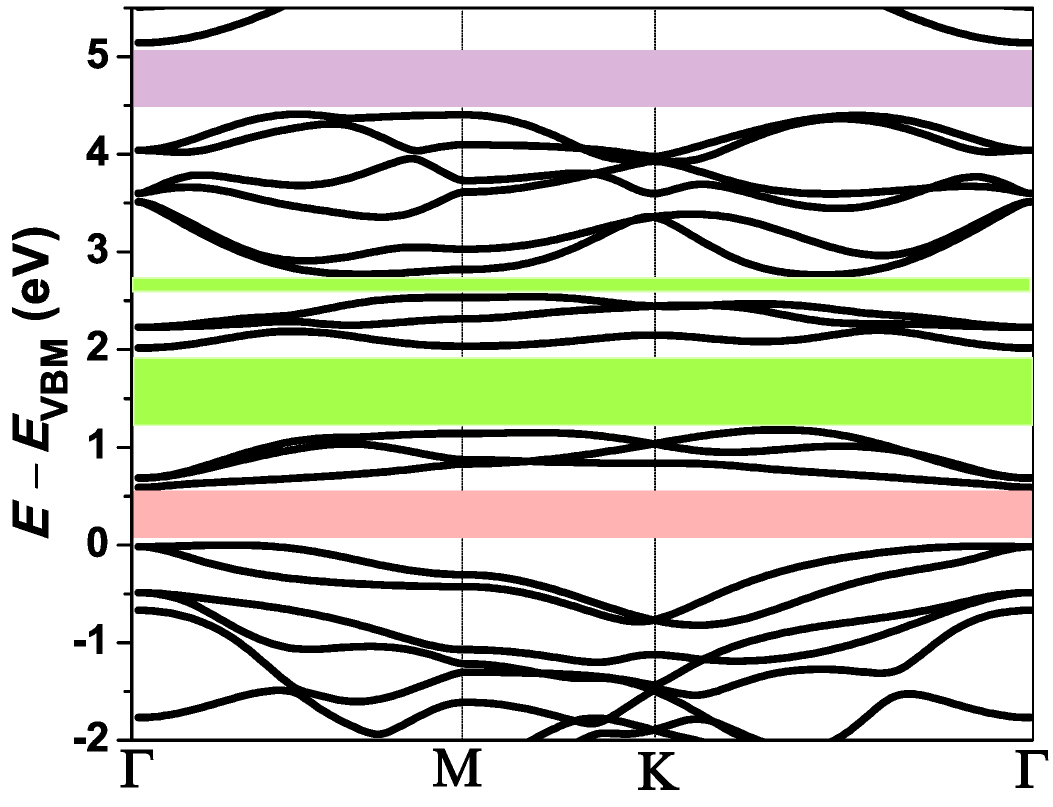}};
          \node [anchor=north west] (imgF) at (0.52\linewidth,.26\linewidth){\includegraphics[width=0.33\linewidth]{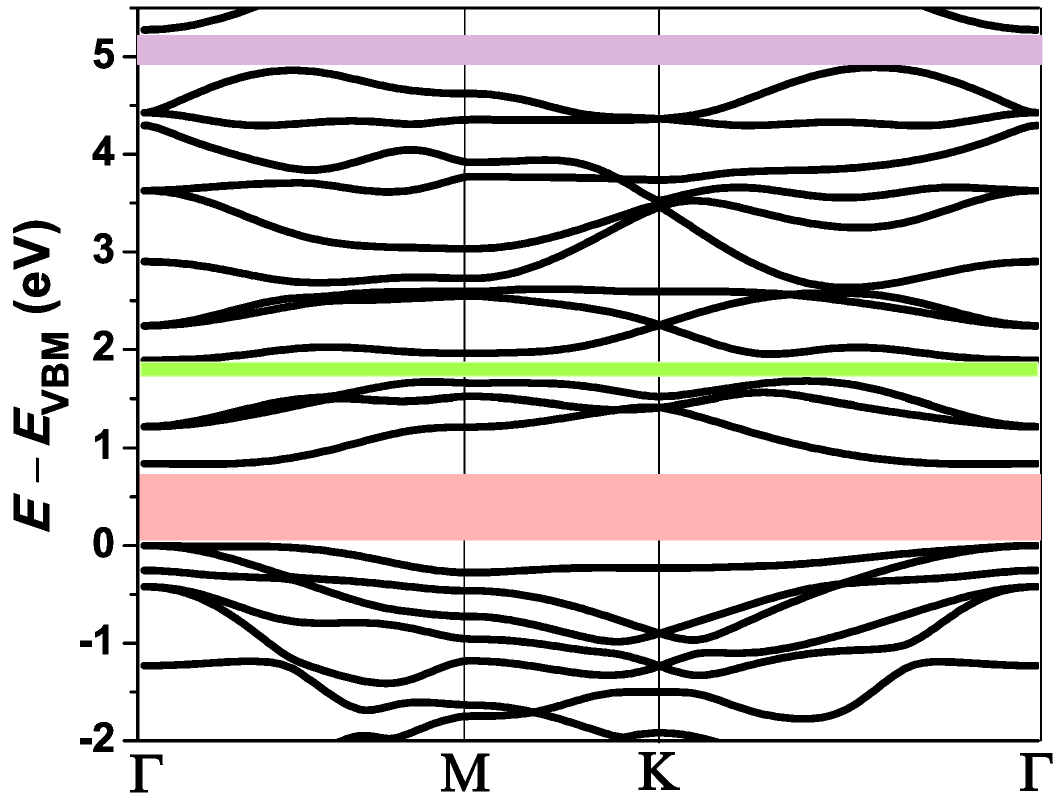}};
          
            \draw [anchor=north west] (-0.15\linewidth, .61\linewidth) node {(a) {\fontfamily{Arial}\selectfont \textbf{$2\times1$}}};
            \draw [anchor=north west] (0.19\linewidth, .61\linewidth) node {(b) {\fontfamily{Arial}\selectfont \textbf{$2\times2$}}};
           \draw [anchor=north west] (0.52\linewidth, .61\linewidth) node {(c) {\fontfamily{Arial}\selectfont \textbf{$4\times4$}}};
            \draw [anchor=north west] (-0.15\linewidth, .30\linewidth) node {(d) {\fontfamily{Arial}\selectfont \textbf{$3\times3$}}};
            \draw [anchor=north west] (0.19\linewidth, .30\linewidth) node {(e) {\fontfamily{Arial}\selectfont \textbf{$\sqrt{3} \times \sqrt{3}$ 2B}}};
           \draw [anchor=north west] (0.52\linewidth, .30\linewidth) node {(f) {\fontfamily{Arial}\selectfont \textbf{$\sqrt{3} \times \sqrt{3}$ 4B}}};          
            \end{tikzpicture}
			}
			
			\caption{Electronic bandstructure of pristine reconstructed structures, in the Brillouin zone of each structure's primitive cell. The colored stripes highlight the gaps, with pink as the fundamental band gap, purple as a gap common to all structures due to transition between Mo $s$ and $d$-orbitals, and green for gaps in between those two. The dotted line in a) shows the small band overlap at this energy.}
		\label{fig:Band_structure}
			
		\end{figure}

	We found a variety of electronic structures in pristine and Ni-doped reconstructions. For pristine phases, band gaps are shown in Table \ref{tab:1T_structure_properties}, density of states (DOS) in Fig. S6, and bandstructures in Fig. \ref{fig:Band_structure}. For Ni-doped phases, the band gaps are in Table S1, DOS in Fig. S7 and bandstructures in Fig. S9. The pristine 1T phase is metallic \cite{lattice,T_3R_bandgap}, but becomes semiconducting with most distortions due to a Jahn-Teller mechanism.\cite{1T_phase_review} This occurred with doping in most cases we considered; only Mo-substituted $2\times2$ and $3\times3$,  S-substituted $4\times4$, and S-atop  $2\times2$ and $4\times4$ remained metallic. %
	All of the pristine reconstructions obtained from the doped ones are semiconducting,
	albeit with a very small gap in $2\times1$, as in \citet{1T_prime_band} We calculated the $\mathbb{Z}_2$ topological index of the pristine and doped reconstructions using Z2Pack,\cite{Z2pack1,Z2pack2} but found them to be topologically trivial, other than $2\times1$\cite{1T_topological_insulator}. We did however notice other unusual features in the bandstructures of the pristine phases: there are multiple gaps in the conduction bands (Table \ref{tab:1T_structure_properties}(b), Fig. \ref{fig:Band_structure}).
	Undistorted 1T has two conduction-band gaps even though it is metallic (Fig. S8a). Only rarely have bandstructures been published extending high enough in energy to see these gaps,\cite{1T_bandgap_variation} and they do not seem to have been discussed in the literature. These gaps are preserved or increase in number in distorted phases (except $2\times1$). DOS plots (Figs. S6-S7) confirm the presence of these gaps. We note that while in general the fundamental gap is underestimated by PBE, other aspects of the bandstructure are typically well estimated. We verified that PBE+$U$ and spin-orbit coupling do not significantly change the bandstructure (e.g. Fig. S11). In all calculated structures, a gap of 0.3--0.7 eV exists $\sim 5$ eV above the VBM, between bands of mostly Mo $d$ character below and Mo $s$ character above (Fig. S8b). A similar gap $\sim 5$ eV above the VBM is found in ReS$_2$ bilayers, between Re $d$ and Re $s$ bands, as probed by time--resolved second harmonic generation.\cite{second_gap} Similar second gaps are reported in other metal chalcogenides.\cite{GaSe,Cu3SbS4} However, the other gaps found in 1T distorted phases are quite unusual, and there does not seem to be a standard explanation for such gaps.
	
	The origin of these gaps is complex. The gaps are induced by perturbations to the potential felt by the electrons at the reconstructed supercell periodicity, as for mini-bands in superlattices,\cite{minibands} with the perturbation provided by Mo clustering. What explains the patterns in different structures?
	The gaps are not due to a transition between different types of orbitals as they are within bands all due to Mo $d$ and S $p$-orbitals. We examined the wavefunctions of band complexes between gaps, and we find some common patterns. Some gaps are transitions between bands localized on clustered and non-clustered Mo orbitals. In $\sqrt{3} \times \sqrt{3}$ 2B and 4B, we see H\"uckel-type patterns of bonding, non-bonding, and anti-bonding among $d$-orbitals in a band complex. We examined local symmetries of Mo atoms for crystal field splitting (Table S3), and find mostly symmetry elements ($C_1, C_{2v}$) that break the degeneracy of $d_{xz}$ and  $d_{yz}$, and $d_{x^2-y^2}$ and $d_{xy}$. This breaking of degeneracy contributes to gap openings, and some of the gaps are found to be transitions between different crystal-field-split $d$-orbital shapes. In the simplest case of undistorted 1T, the gap around 2 eV above the Fermi energy has mostly $d_{z^2}$ character below and mostly $d_{xz}$ and $d_{yz}$ above (Fig. S8b). We do not see any clear correlation between the multiple gap patterns and the overall symmetry, number of inequivalent Mo atoms, or symmetry of the Mo sublattice. In a tight-binding picture, loss of symmetry can increase the variation of on-site energies; for small hoppings (and bandwidths), this could open up gaps, but for large hoppings, it could overlap bands and close gaps. We investigated the L\"owdin charges on Mo atoms and found significant variation (Fig. S12), which would correlate with on-site energies, but there is no clear correlation of the charges with number or magnitude of gaps. Ultimately we conclude that the gaps, and the number of bands between them, are due to a complex interplay of reconstructed periodicity, Mo cluster bonding, crystal-field splitting, and overall symmetry. Wannier-function-based analysis \cite{PasquierPRB} may shed further light on the mechanisms.

	These multiple gaps could possible be utilized in intermediate band solar cells (IBSCs) \cite{intermediate_solar}, a third-generation photovoltaic concept which can exceed the Shockley-Queisser limit by allowing the voltage to be increased without losing absorption. In IBSCs, electrons can be excited from the valence band across a first band gap to an intermediate band (from which recombination is slow), and then excited again across a second band gap. Whereas in a standard solar cell the single band gap controls both the voltage and the absorption, here the resulting voltage is related to the sum of the two band gaps. Such intermediate bands for solar cells are typically due to dopants,\cite{multiple_gaps, osti} which can however cause undesired scattering and recombination. In our case, however, we have multiple gaps even in the pristine phases due to reconstruction. Ideally, an IBSC should have a partially filled intermediate band, to facilitate electronic transitions,\cite{metallic_bandstr}, which could be obtained by electrostatic gating, or realized by some of the Ni-doped structures (Fig. S9 (b,d)) that have metallic intermediate bands.
	Multiple gaps, in particular in the metallic cases or with gating, could also be applied for transparent conducting materials (TCMs)\cite{TCO} -- in some successful TCMs, the bandstructure has a partially filled conduction band, providing conductivity, with a large second gap above it, providing transparency. As with IBSCs, such bandstructures have typically been provided by doping, but here could be realized in an undoped system which may have better mobility. A final application is nonlinear optics, as studied for ReS$_2$,\cite{second_gap} since multiple gaps enable multiple simultaneous resonances for sum- or difference-frequency generation. Phases studied here are noncentrosymmetric and therefore have nonzero $\chi^{(2)}$, as has been in shown for $\sqrt{3}\times\sqrt{3}$ 4B\cite{1T_prime_3_exp} and $2\times2$\cite{polarization_exp}.
	
	We find spontaneous polarization in the reconstructions. Pristine $\sqrt{3} \times \sqrt{3}$ 4B has been calculated to be ferroelectric with an equivalent bulk polarization of 0.18 $\mu$C/cm$^2$ (using the S-S distance as the height along the $z$-direction)\cite{1T_ferroelectric}. Recently the out-of-plane polarization for a bulk $2\times2$ with layer height of 5.96 \AA\ was calculated\cite{polarization_exp} to be 0.04 $\mu$C/cm$^2$ (we obtain 0.06 $\mu$C/cm$^2$). We will focus on the polarization per unit area as befitting a monolayer 2D material. We calculate polarization magnitudes of: $2\times2$, $4.06 \times 10^{-4}\  e$/\AA; $\sqrt{3} \times \sqrt{3}$ 4B, $7.18 \times 10^{-4}\ e$/\AA\  (consistent with \citet{1T_ferroelectric}); $\sqrt{3} \times \sqrt{3}$ 2B, $2.19 \times 10^{-3}\ e$/\AA\ ($3\times$ larger than 4B); $4\times4$, $1.34 \times 10^{-2}\ e$/\AA; and $3\times3$, $5.37 \times 10^{-2}\ e$/\AA. The polarizations are in the $-z$ direction for our $\sqrt{3} \times \sqrt{3}$ 2B and $\sqrt{3} \times \sqrt{3}$ 4B structures, as in-plane polarization is forbidden in $C_{3v}$. There is in-plane polarization for $2\times2$ ($C_s$), $3\times3$ ($C_1$) and $4\times4$ ($C_1$), with polarization vectors $\left( 3.45, 0, 2.15 \right)\times10^{-4}\ e$/\AA, $\left( 380, -380, -1.76 \right)\times10^{-4}\ e$/\AA, and $\left( 126, -44.3, -3.07 \right)\times10^{-4}\ e$/\AA, respectively. These $3\times3$ and $4\times4$ structures' polarizations are 132 and 33 times larger than $2\times2$. Using the bulk $2\times2$ layer height, these correspond to bulk polarizations of $\sim$8 $\mu$C/cm$^2$ and $\sim$2 $\mu$C/cm$^2$ respectively, only one order of magnitude smaller than standard ferroelectrics PbTiO$_3$ (79 $\mu$C/cm$^2$) and  PbZrO$_3$ (70 $\mu$C/cm$^2$).\cite{ferroelectric_barrier} For comparison, $2\times1$ has zero polarization due to $C_{2h}$ symmetry. We can make a simple estimate of ferroelectric switching, considering coherent reversal via the (paraelectric) undistorted 1T structure\cite{ferroelectric_barrier}, and using the energy differences in Table \ref{tab:1T_structure_properties}. We obtain barriers of 0.03--0.09 eV/atom, comparable to PbTiO$_3$ ($\sim$0.01 eV/atom) and PbZrO$_3$ ($\sim$0.04 eV/atom).\cite{ferroelectric_barrier} Moreover, there could be a lower-energy switching pathway. Out-of-plane polarization switching was recently observed in an experiment\cite{polarization_exp} for the $2\times2$ structure, for which we estimate a barrier of 0.21 eV. Since all of the structures have similar or lower barriers, these results suggest the polarization is switchable in those cases too. The experimental results suggest that monolayer ferroelectric properties can be retained when stacked into a multilayers, for potential nanoelectronic devices.

        We investigated magnetism in doped structures, given its presence in some structures of doped 1H-MoS$_2$ \cite{Hub_U_Mo, Karkee2020}.
        We found 3 ferromagnetic cases and 1 antiferromagnetic case, with moments on the Ni site (Table \ref{tab:magnetization}, Table S2). We estimated Curie temperatures with mean-field theory (SI Sec. 1), and find high $T_C$ for the metals. Metals cannot have polarization in a periodic direction, but two of these cases have out-of-plane polarization, making them ferromagnetic polar metals.\cite{polar_metals} Note that $2 \times 2$ was calculated and measured to be a polar metal as well \cite{polarization_exp} but only in the bulk, not monolayers. We also have unusual ferromagnetic and antiferromagnetic semimetals \cite{EuB6_Zhang,Tang},  
        which have both out-of-plane and in-plane polarization, making them (anti)ferromagnetic-ferroelectric multiferroics, rarely reported in 2D.\cite{multiferroics,Lyu_multiferroic} Polarization for all doped structures is in Table S1. The in-plane polarization is switchable via undistorted 1T where it is zero by symmetry, with switching energies per atom comparable to the pristine phases, and there could be other lower-barrier pathways.
        Doped structures (except for Mo substitution) have out-of-plane polarization due mainly to Ni (with oxidation state $2-$, SI Sec. 1) rather than distortion, which therefore cannot be feasibly switched. Mo substitution has switchable (though small) polarization since the undistorted 1T is centrosymmetric. The other cases are an unusual kind of improper ferroelectric controlled by in-plane distortion \cite{improper_ferroelectric}, where only the in-plane polarization (up to $\sim 10^{-3}\ e/$\AA) is switchable. We note this type is distinct from recently reported ferroelectric doped 2D materials, which switch via interlayer sliding rather than intralayer distortions.\cite{Sui_doped_ferroelectric} 

 \begin{table}[htbp]
 
 \centering
 \begin{tabular}{|c|r|r|r|c|r|}
 \hline
 Doping type & $\mu$ ($\mu_B$) & $\Delta E_{\rm mag}$ (meV) & $T_C$ (K) & Type & Polarization ($e$/\AA) \\
 \hline \hline
 $2\times2$ S subs. & 0.05   & -0.19 & 19.6 & semimetal & $(-1.06, 1.18, -3.06) \times 10^{-3}$ \\
 \hline
 $4\times4$ S subs. & 0.78  & -27.91 & 493 & metal & $(0, 0, -6.60) \times 10^{-4}$ \\
 \hline
 $2\times2$ S-atop &  0.30 &  -17.32 & 485 & metal & $(0, 0, -1.14) \times 10^{-2}$\\
 \hline
 $3\times3$ S-atop & 0 [0.07] & -0.09 & - & semimetal & $(6.08,-6.08, -49.3) \times 10^{-4}$ \\
 \hline
 
 \end{tabular}
 \caption{ Magnetic doped structures.\textsuperscript{\emph{a}}}
 \textsuperscript{\emph{a}} Moment $\mu$ per Ni atom and energy difference $\Delta E_{\rm mag}$ per Ni atom of spin-polarized state \textit{vs.} the non-spin-polarized paramagnetic state. Moment in square brackets for $3\times3$ S-atop is the integrated absolute magnetization, indicating weak antiferromagnetism.
 \label{tab:magnetization}
 \end{table}

In conclusion, we found that Ni--doping 1T MoS$_2$ induces a panoply of different reconstructions with Mo clustering, due to the effect of Ni on neighboring bonds. Removing or replacing the dopant and relaxing (and adding a small perturbation in some cases) leads to pristine reconstructions are dynamically stable. These resulting phases include not only the  typically reported ones but also $3 \times 3$ and $4 \times 4$. Two versions of $\sqrt{3} \times \sqrt{3}$ with two and four Mo-Mo bonds are present initially, but only the latter is dynamically stable.
These structures show large ferroelectric polarizations and low switching barriers. Most notably, compared to the experimentally measured $2 \times 2$\cite{polarization_exp}, $3 \times 3$ has polarization higher by two orders of magnitude, is more stable, and has a significantly larger band gap, which reduces difficulties with charge leakage and conductivity. Most Ni-doped structures are semiconductors but they also include unusual ferromagnetic polar metals, multiferroic semimetals, and in-plane improper ferroelectrics. Pristine and doped phases showed multiple gaps in the conduction bands, a phenomenon which has potential application in intermediate band solar cells, transparent conducting materials, and nonlinear optics. Our calculations suggest that Ni-doping of 1T-MoS$_2$ (perhaps including intercalation) could be a way to synthesize a range of distorted pristine phases, by removing or replacing dopants through methods such as dissolution or evaporation of adatoms or intercalants,\cite{1T_prime_Kdoping} or ion exchange for Mo or S substitution \cite{Chen_ionexchange,nanoflowers}. Other possible techniques are
epitaxy on a Ni-doped monolayer or other suitable substrate\cite{CHAE201965}; or stabilization with appropriate ligands.\cite{ligands} Experiments have shown that 1T phases can be stable even under reaction conditions of ion exchange \cite{nanoflowers} or etching of S atoms.\cite{1T_triple_prime_exp} Such structures, including the potential analogues involving other transition-metal dichalcogenides and transition-metal dopants, are interesting for applications in electronics, optics, energy storage, and catalysis, and may offer other intriguing properties. 

\section*{Computational Methods}
Our calculations use density functional theory (DFT) in the plane-wave pseudopotential code Quantum ESPRESSO \cite{QE_2}. Further details are given in SI Sec. 1. We use the PBE exchange-correlation functional \cite{original_pbe} and Grimme-D2 \cite{Grimme} van der Waals correction for consistency with bulk calculations (SI Sec. 2), except for formation energies with respect to Mo and Ni, for which we use PBE alone.\cite{enrique} Polarization is calculated with the dipole correction method \cite{dipole_corr} out of plane, and Berry-phase method\cite{RevModPhys} in plane. Phonon bandstructures to assess dynamical stability were calculated by density-functional perturbation theory.\cite{Baroni} We work with neutral systems, periodic in plane but separated by vacuum out of plane. 

\section*{Acknowledgments}
We acknowledge Enrique Guerrero for preparing the phase diagram. This work was supported by UC Merced start-up funds and the Merced nAnomaterials Center for Energy and Sensing (MACES), a NASA-funded research and education center, under awards NNX15AQ01 and NNH18ZHA008CMIROG6R. This work used computational resources from the Multi-Environment Computer for Exploration and Discovery (MERCED) cluster at UC Merced, funded by National Science Foundation Grant No. ACI-1429783, and the National Energy Research Scientific Computing Center (NERSC), a U.S. Department of Energy Office of Science User Facility operated under Contract No. DE-AC02-05CH11231.

\section*{Supplementary Information}
Computational details; analysis of formation energies; full structural, electronic, polarization, and magnetic properties of Ni-doped 1T-MoS$_2$; doping formation energies as a function of concentration; doped bandstructures; doped and pristine density of states; and further information on phonon bandstructure and dynamical instabilities (PDF). PBE+Grimme-D2 relaxed coordinates for Ni-doped 1T MoS$_2$ and reconstructed pristine 1T structures, in XCrySDen format (ZIP).

\providecommand{\latin}[1]{#1}
\makeatletter
\providecommand{\doi}
  {\begingroup\let\do\@makeother\dospecials
  \catcode`\{=1 \catcode`\}=2 \doi@aux}
\providecommand{\doi@aux}[1]{\endgroup\texttt{#1}}
\makeatother
\providecommand*\mcitethebibliography{\thebibliography}
\csname @ifundefined\endcsname{endmcitethebibliography}  {\let\endmcitethebibliography\endthebibliography}{}

\end{document}